\RequirePackage{lineno}
\documentclass[aps,prl,twocolumn,superscriptaddress,groupedaddress,preprintnumbers]{revtex4}  
\usepackage{graphicx}  
\usepackage{dcolumn}   
\usepackage{bm}        
\usepackage{amssymb}   
\usepackage{multirow}
\usepackage{epsfig}
\usepackage{amsmath}
\usepackage[colorlinks, linkcolor=blue]{hyperref}
\usepackage{ulem}
\usepackage{float}
\usepackage{color}
\usepackage{subcaption}
\usepackage{caption}

\begin{document}

\modulolinenumbers[3]
\lefthyphenmin=2
\righthyphenmin=2

\widetext

\title{Flavour asymmetry of antiquarks in nucleon and nucleus}
\author{WenHao Ma} \affiliation{Department of Modern Physics, University of Science and Technology of China, Jinzhai Road 96, Hefei, Anhui 230026, China}
\author{Siqi Yang}\email{yangsq@ustc.edu.cn} \affiliation{Department of Modern Physics, University of Science and Technology of China, Jinzhai Road 96, Hefei, Anhui 230026, China}
\author{MingZhe Xie} \affiliation{Department of Modern Physics, University of Science and Technology of China, Jinzhai Road 96, Hefei, Anhui 230026, China}
\author{Minghui Liu} \affiliation{Department of Modern Physics, University of Science and Technology of China, Jinzhai Road 96, Hefei, Anhui 230026, China}
\author{Liang Han} \affiliation{Department of Modern Physics, University of Science and Technology of China, Jinzhai Road 96, Hefei, Anhui 230026, China}
\author{C.-P. Yuan} \affiliation{Department of Physics and Astronomy, Michigan State University, East Lansing, MI 48823, USA}

\begin{abstract}
Over the years, comprehensive experiments have shown a fact that the nucleons, such as the proton and 
neutron, are formed by 
not only the ``valence'' up and down quarks which were thought to comprise the nucleons in a 
simple constituent picture, 
but also ``sea'' quarks which can be any other flavour. However, it is still unknown how sea 
quarks are generated inside the nucleons. Since 1990s, measurements on high energy deuterons (formed by a 
proton and a neutron) 
indicated that the anti-down quark contribution was higher than the anti-up quark in the proton, based on 
the assumptions of the proton-neutron isospin symmetry and 
a small nuclear effect of the deuteron.
Henceforth, sea quarks are considered to be generated via some flavour-asymmetrical
mechanisms. Here we report an analysis on a series of new 
measurements from pure proton interactions which are free from those assumptions,  
unexpectedly showing that the anti-down quark component is rather consistent with the anti-up quark. 
It appears to be evidence that the previously observed asymmetry was caused by 
an unknown nuclear effect in the deuteron, rather than by a difference between 
antiquarks. We anticipate this work to be an essential new discovery and 
a motivation for studying nuclear structure, both experimentally and theoretically, 
at high energy scales, as it now appears fundamentally different from our 
understanding established in the past.
\end{abstract}
\maketitle

\section{\bf Introduction and conclusion}

In the simple picture of the constituent model established about sixty years ago, 
the proton, as one important nucleon building up the material world, 
is comprised of two up ($u$) quarks and one down ($d$) quark. They are bonded by 
the strong forces described by the quantum chromodynamics (QCD). 
However, when the energy of the proton-involved interactions is high enough that 
the protons are believed to break into pieces, the fragments knocked out from the protons are 
found to be not only the ``valence'' $u$ and $d$ quarks, but also ``sea'' quarks which can be any other flavour, 
including even antiquarks. The fundamental mechanism of how sea quarks generate 
inside the proton remains as a puzzle. 
At beginning, the anti-down ($\bar{d}$) and anti-up ($\bar{u}$) quarks 
were expected to have similar distributions in the proton, 
as the perturbative QCD predicts a symmetrical generation of $\bar{d}$ and $\bar{u}$ 
from the gluon splitting at 
high energy scale where the difference between their masses can be ignored. Unexpectedly, 
the distribution of $\bar{d}(x)$ as a function of the momentum fraction $x$ 
was found to be significantly higher than $\bar{u}(x)$ from a 
series of experiments since 1990s~\cite{NuSea, SeaQuest, NA51, NMC, HERMES}, known 
as the SU(2) flavour asymmetry of the light quark sea. 
These data provide strong evidence for a non-symmetrical sea quark generation 
mechanism and have become the most important constraints on antiquarks in proton 
structure studies~\cite{CT18, MSHT, NNPDF}. 
Based on these measurements, 
various models have been developed, trying to explain the asymmetry as a non-perturbative 
effect~\cite{mesoncloud, chiral1, chiral2, chiral3, soliton1, soliton2, soliton3}. 

Although flavour asymmetry pertains specifically to sea quarks in the proton, directly 
measuring $\bar{d}(x)$ and $\bar{u}(x)$ from proton interactions has proven rather difficult. 
This is because these antiquarks always mix in the initial state, producing similar 
final states that are indistinguishable.
Therefore, all those 
measurements mentioned above were performed with deuterons. 
For example, in Fermilab's NuSea 
and SeaQuest experiments~\cite{NuSea, SeaQuest}, and CERN's NA51~\cite{NA51} experiment, 
protons are accelerated and directed to hit hydrogen and deuteron targets. 
The cross sections of the 
proton-to-hydrogen ($\sigma_{pH}$) and proton-to-deuteron 
($\sigma_{pD}$) interactions are measured and compared with each other. 
Based on the assumptions of a small nuclear effect that a deuteron can be treated simply as the sum 
of a proton ($p$) and a neutron ($n$), and the isospin symmetry that the quark distributions in 
the proton and the neutron can be related as $u_p=d_n$, $d_p=u_n$, $\bar{u}_p=\bar{d}_n$, and 
$\bar{d}_p=\bar{u}_n$, the ratio of the antiquark distributions can be 
measured as $\bar{d}/\bar{u} \approx \sigma_{pD}/\sigma_{pH} - 1$. Other measurements, 
such as NMC~\cite{NMC} and 
HERMES~\cite{HERMES}, were based on lepton beams instead of proton beam, but still used deuteron target. 
Thus, they also assumed a small nuclear effect of deuteron. 
These assumptions are widely taken at low energy scales, where nuclei are 
nearly stationary, and are also used in nucleon structure studies at high energy scales. 
However, nuclear structure may differ from the understanding 
established in low-energy scenarios: at high energies, relativistic effects can 
spatially compress a nucleus, introducing unknown impacts on its observed 
structure. As will be discussed in detail later, this high-energy nuclear structure has 
never been experimentally verified.  

In this article, we report a determination of the $\bar{d}(x)$-to-$\bar{u}(x)$ ratio at $x\sim 0.1$, 
using experimental 
data from pure proton interactions. This is free from the assumptions taken in the deuteron measurements. 
The major constraint on $\bar{d}/\bar{u}$ comes from a series of novel measurements at the proton-(anti)proton 
colliders in recent years, which can decouple the contributions of different quark flavours and reveal 
the information of their relative ratio~\cite{AFBFactorization, D0PuPd, CMSPuPd}. 
It is achieved from the Drell-Yan process of $hh(q\bar{q})\rightarrow Z/\gamma^* \rightarrow \ell^+\ell^-$, 
where a pair of quark and antiquark arise from the hadron collision ($hh$), 
and annihilate into a lepton pair ($\ell^+\ell^-$). 
The Drell-Yan process is famous for its spatial symmetry breaking at the pole region 
of the $Z$ boson mass around 90 GeV, which is a unique feature of the weak force. 
Particularly, the cross section of the forward events, which are defined as the outgoing lepton ($\ell^-$) 
directing to the hemisphere same as the incoming quark ($q$), 
is different from that of the 
backward events, defined as $\ell^-$ pointing to the opposite direction of $q$. 
Such forward-backward asymmetry depends on the flavour of the quarks 
coupled with the $Z$ boson, and can be precisely predicted in the electroweak theory, 
thus can be used to separate the contributions from the up and down (anti-)quarks.
As discussed in Ref.~\cite{AFBFactorization},  
the following ratios can be extracted from the forward-backward asymmetry in the 
proton-proton collisions and proton-antiproton collisions, respectively:

\begin{footnotesize}
\begin{eqnarray}\label{eq:ratio}
  R_{pp} &=& \frac{d(x_1)\bar{d}(x_2) - \bar{d}(x_1)d(x_2)}{u(x_1)\bar{u}(x_2) - \bar{u}(x_1)u(x_2)} \times \mathcal{N}_\text{EW}\nonumber \\
  R_{p\bar{p}} &=&  \frac{d(x_1)d(x_2)-\bar{d}(x_1)\bar{d}(x_2)}{u(x_1)u(x_2) - \bar{u}(x_1)\bar{u}(x_2)} \times \mathcal{N}_\text{EW}
\end{eqnarray}
\end{footnotesize}

\noindent where $x_1>x_2$ are the momentum fractions of the two quarks in the initial state. $\mathcal{N}_\text{EW}$ 
is a normalization factor contributed by the hard scattering processes involving $q-\bar{q}-Z/\gamma^*$ couplings, 
which is precisely known from the higher order calculations of the electroweak theory.
Contributions from other quarks, 
such as the strange ($s\bar{s}$), charm ($c\bar{c}$) and bottom ($b\bar{b}$), 
are nearly cancelled, because the distributions of these quarks 
are almost equal to their corresponding antiquarks ($s(x)\approx \bar{s}(x)$, $c(x)\approx \bar{c}(x)$, 
$b(x)\approx \bar{b}(x)$)~\cite{cancellation}, so that no ``spatial direction'' can be defined. 

The $R$ parameter has been subsequently measured by the D0 collaboration 
at the Fermilab's Tevatron~\cite{D0PuPd}, which is a proton-antiproton collider with 1.96 TeV 
colliding energy, 
and from the 
proton-proton data collected by the CMS detector at the CERN's Large Hadron Collider (LHC) at 
8 TeV~\cite{CMSPuPd}. 
Due to the high beam energy, the dilepton system of the Drell-Yan process is highly boosted, 
giving $x_1$ around 0.1, and $x_2$ of $\mathcal{O}(0.01)$ or even smaller. At such a small 
$x_2$ region, distributions of light quarks are dominated by the perturbative QCD 
contributions, so that are comparable. Therefore, $R_{pp}$ and $R_{p\bar{p}}$ 
can be approximately written as:

\begin{footnotesize}
\begin{eqnarray}
R_{pp,p\bar{p}} &\approx& \frac{d(x_1) - \bar{d}(x_1)}{u(x_1)-\bar{u}(x_1)}\times \mathcal{N}_\text{EW} \nonumber \\
  &=& \frac{d(x_1)+\bar{d}(x_1) - 2\bar{d}(x_1)}{u(x_1) +\bar{u}(x_1) - 2\bar{u}(x_1)} \times \mathcal{N}_\text{EW},
\end{eqnarray}
\end{footnotesize}

\noindent where 
$u(x_2)$, $\bar{u}(x_2)$, $d(x_2)$ and $\bar{d}(x_2)$ are roughly cancelled in the ratio. 
$u+\bar{u}$ and $d+\bar{d}$ can be sufficiently constrained by comprehensive measurements 
of the electron-proton deep inelastic scattering (DIS) experiments, such as HERA~\cite{HERA}. 
Thereupon, the antiquark ratio of $\bar{d}(x)/\bar{u}(x)$ can be determined by analyzing the 
observed $R$ parameter together with a group of previously performed measurements of proton-interactions, 
including lepton-proton DIS data from HERA and BCDMS,  
the proton-proton collision data 
from the CERN's Large Hadron Collider (LHC), the proton-antiproton collision data from the 
Fermilab's Tevatron, and the proton-hydrogen target data from the Fermilab's E866 (NuSea). 
These data, as summarized in Ref.~\cite{postCT18}, are selected from the CTEQ-TEA global analysis 
of the proton structure, with all the nucleus-based experiments excluded. 
The only exception is the neutrino-iron DIS data from the NuTeV collaboration~\cite{NuTeV}, which 
specifically constrains the strange quark distribution. We include this data to avoid potential 
bias in the $\bar{d}(x)/\bar{u}(x)$ ratio that could arise from an unconstrained $s(x)$ distribution. We 
have verified that including or excluding the NuTeV dataset produces only a minor numerical 
difference in the $\bar{d}(x)/\bar{u}(x)$ ratio.
These previously measured data do not provide direct constraint 
on $\bar{d}/\bar{u}$, but they are important for accounting for other quark and gluon contributions and 
higher order QCD corrections. 
The strategy of this analysis follows the CT18 global analysis of the 
proton structure~\cite{CT18}(but not exactly the same), 
in which the distributions of 
quarks are factorized and determined using the Hessian method~\cite{Hessian}. 
To fully account for the proton's valence quark number sum rule and 
momentum sum rule, we fit the distributions of all quark flavours across the entire 
momentum fraction range ($x$ in [0, 1]) using the data,  
although this work focuses on the $\bar{d}(x)/\bar{u}(x)$ ratio at $x\sim 0.1$. 
Details of the analysis procedure, which has been specifically modified for this study 
to differ slightly from the CT18 strategy, will be presented 
in the next section.

\begin{figure*}[!h]
\begin{center}
\includegraphics[width=0.6\textwidth]{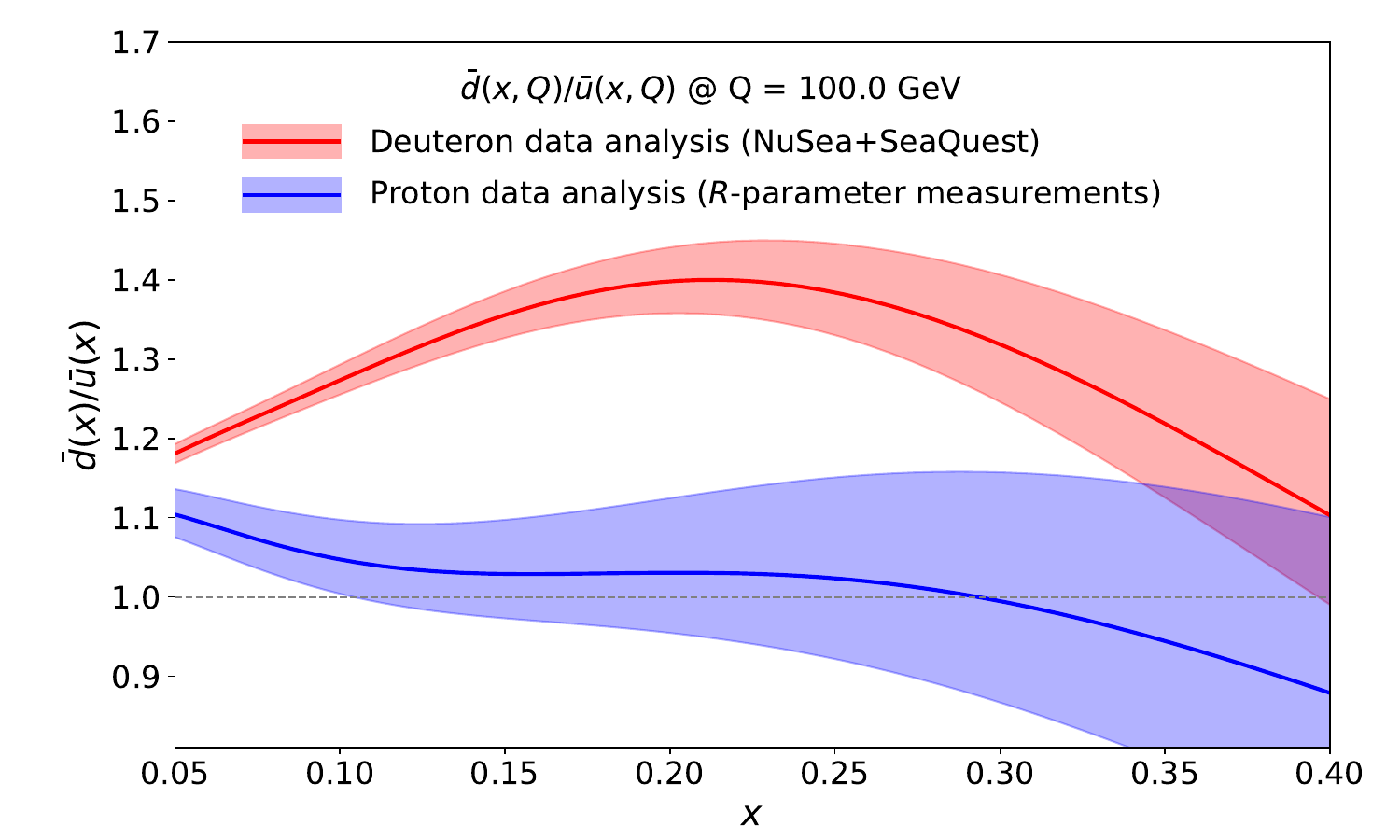}
\caption{\small {\bf Distributions of $\bar{d}(x)/\bar{u}(x)$.} Distributions of $\bar{d}(x)/\bar{u}(x)$, 
obtained from two separated analyses, one using 
pure proton data (blue curve with uncertainty band in same colour) 
and another using deuteron data (red curve with uncertainty band in same colour). 
The proton data-based analysis incorporates experimental results from HERA, BCDMS, LHC, Tevatron, 
and Fermilab's E866 (specifically, only the proton-hydrogen measurement). Additionally, measurements used to 
determine the $R$-parameter are included to constrain the antiquark distributions. 
In contrast, the deuteron-based analysis replaces the $R$-parameter measurements with 
experimental results from the NuSea and SeaQuest collaborations. 
For consistency, all distributions are calculated at the energy scale 
of 100 GeV. The uncertainties of these distributions are estimated by taking into account both the 
statistical errors and experimental systematics of the measurements.}
\label{fig:flavorasym}
\end{center}
\end{figure*}

\begin{figure*}[!h]
\begin{center}
\begin{subfigure}{0.45\textwidth}
\includegraphics[width=\textwidth]{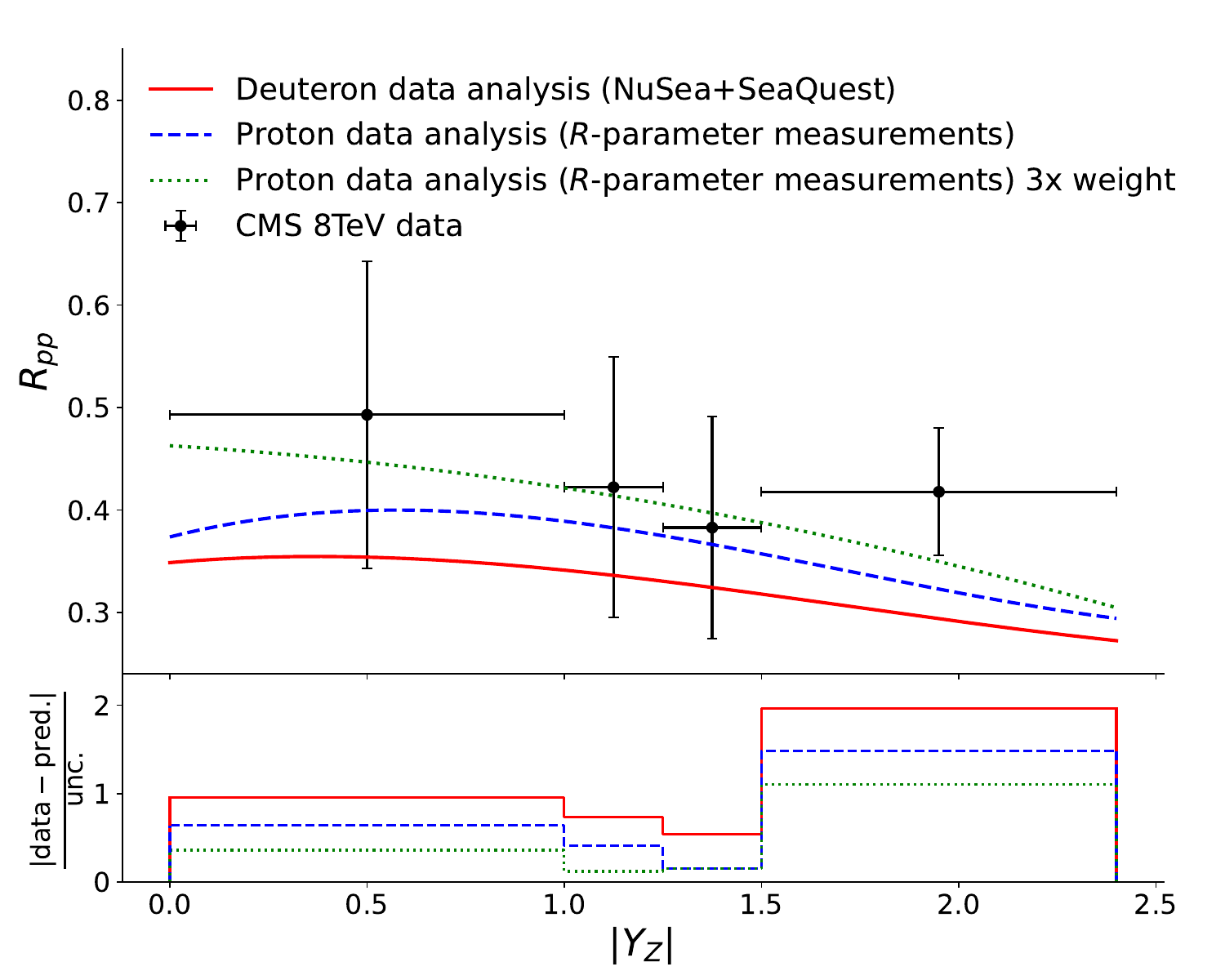}
\caption{}
\end{subfigure}
\begin{subfigure}{0.45\textwidth}
\includegraphics[width=\textwidth]{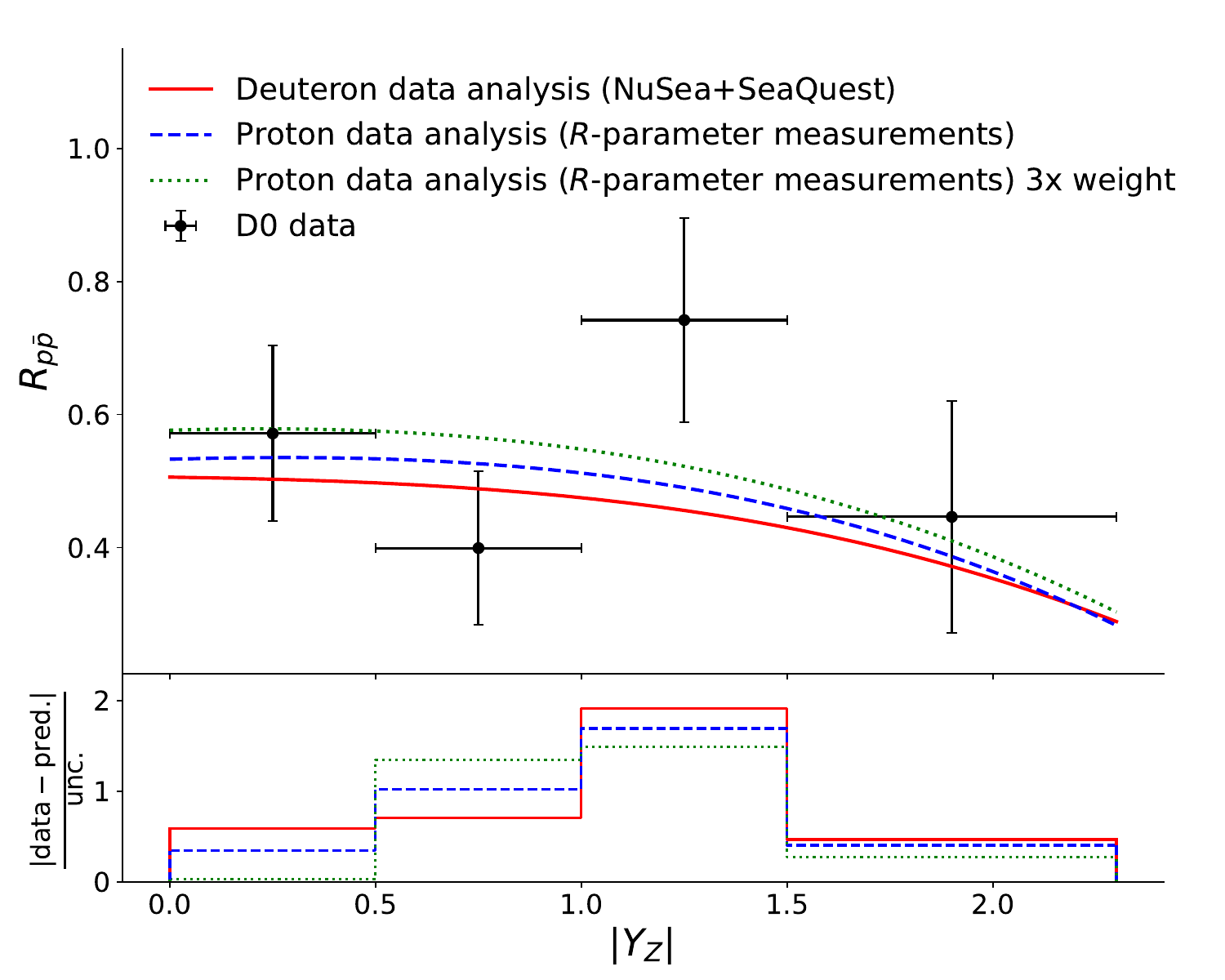}
\caption{}
\end{subfigure}
\begin{subfigure}{0.45\textwidth}
\includegraphics[width=\textwidth]{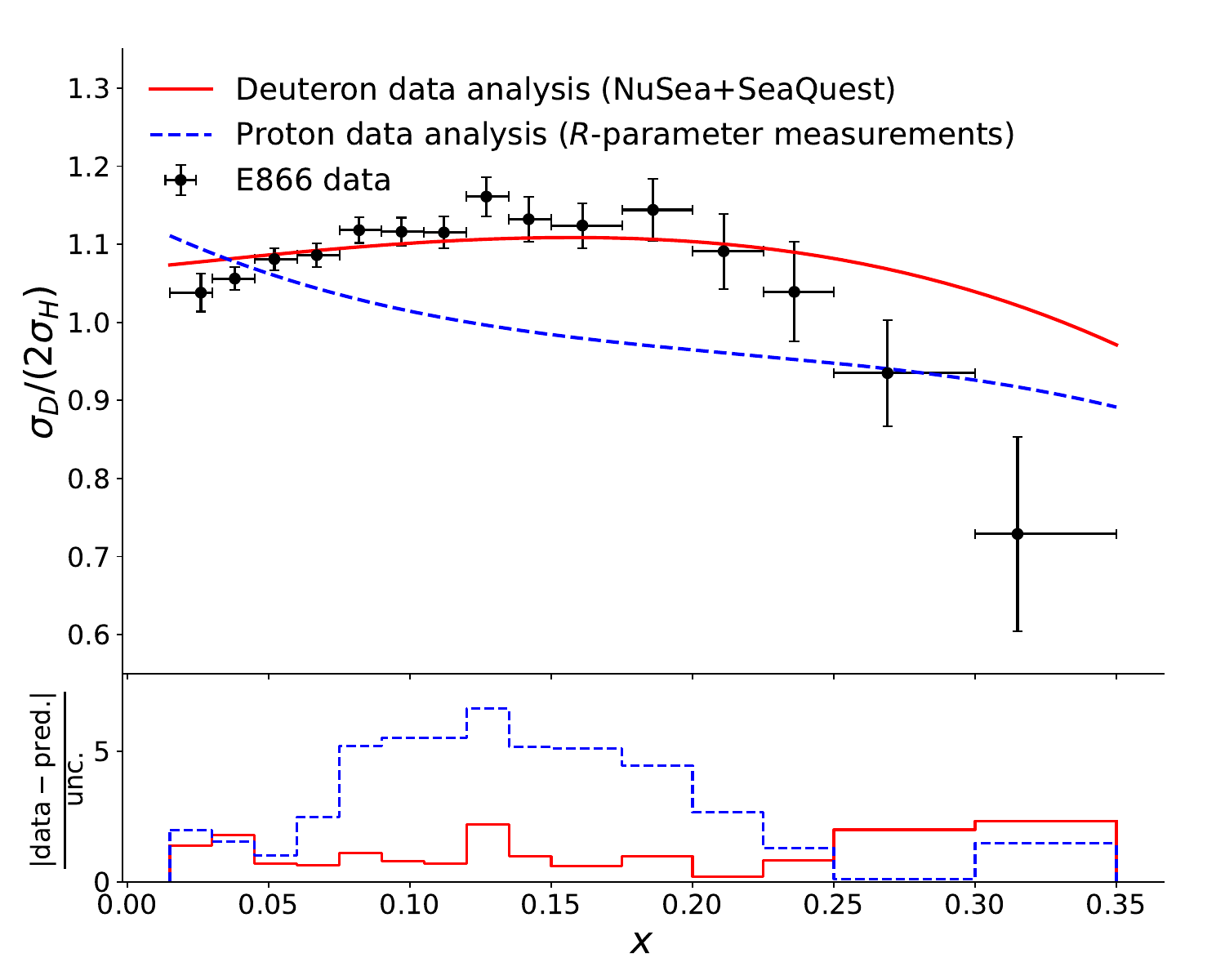}
\caption{}
\end{subfigure}
\begin{subfigure}{0.45\textwidth}
\includegraphics[width=\textwidth]{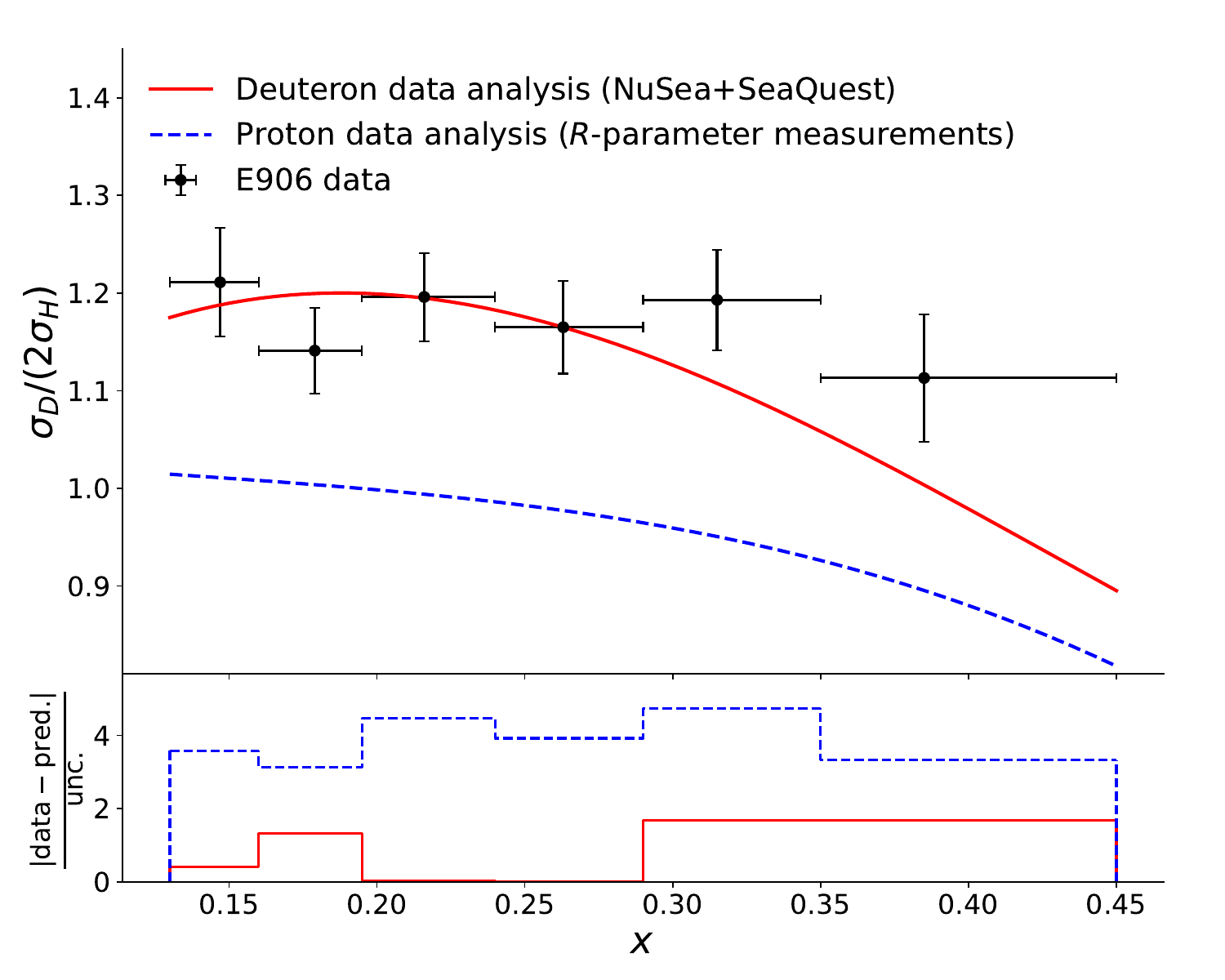}
\caption{}
\end{subfigure}
\caption{\small {\bf Comparison between the data and predictions.} The measured 
$R$ parameters from the CMS (a) and D0 (b) data, together with the measured $\sigma(pD)/[2\sigma(pH)]$ from 
the NuSea (c) and SeaQuest (d) data, are shown. 
The vertical error bars represent the total uncertainties, including both statistical fluctuations and experimental systematics. 
The corresponding theoretical predictions (in solid or dashed curves) 
for these observables are provided for comparison. 
The lower panel in each subfigure displays the uncertainty-normalized difference between the 
predictions and the measurements, i.e., the difference divided by the standard deviation. 
The $R$ parameters are presented as a function of $|Y_Z|$,  
the rapidity of the Drell-Yan pair, while the ratio of $\sigma(pD)/[2\sigma(pH)]$ is presented as 
a function of $x$.}
\label{fig:data}
\end{center}
\end{figure*}

For comparison, a analogous study is conducted by analyzing deuteron data from NuSea 
and SeaQuest, alongside proton-interaction measurements (without the $R$-parameter 
measurements), using the same methodology. 
This study still assumes isospin symmetry between the proton and neutron, while neglecting 
the deuteron's nuclear effect. 
For simplicity, the NA51~\cite{NA51}, NMC~\cite{NMC} and HERMES~\cite{HERMES} results 
are not used in this study, as they provide similar information and experimental conclusion, 
but are less sensitive. 
The ratios of $\bar{d}(x)/\bar{u}(x)$ as a function of $x$, obtained from individual fits using either the pure proton 
data or the deuteron data, are demonstrated in Fig.~\ref{fig:flavorasym}. As discussed, the NuSea 
and SeaQuest data lead to a $\bar{d}(x)/\bar{u}(x)$ greater than unity, indicating an asymmetry 
in the light quark sea of the proton. However, the pure proton data reveals the antiquark ratio to be close to unity, which 
is very different from the deuteron results. In Fig.~\ref{fig:data},  
predictions from the two analyses are compared with the 
measured values of the original experimental observables, i.e. 
the $R$ parameter and $\sigma(pD)/(2\sigma(pH))$. 
As shown, the predictions are consistent with the corresponding data which are included in the analysis. 
However, deviations emerge when the predictions are compared with the other dataset which are not 
included in the fits. 
As a consistency check, we performed an additional analysis of 
pure proton data by tripling the statistical power of the input $R$-parameter data. 
As shown in the figure, the predictions from this new analysis are more consistent 
with the measured $R$ parameter, indicating that the conclusions of our original 
analysis are not driven by random fluctuations. 
This study is included solely as a sensitivity check, whereas the distributions in Fig.~\ref{fig:flavorasym}, which 
represent the main conclusion of this article, are presented without enhancing the 
data power. 

The result from the pure proton interaction measurements on the antiquark ratio is of importance to the 
nuclear structure study. It indicates that the previous asymmetry 
observed from deuteron measurements could possibly caused by some unknown nuclear effects at the 
corresponding energy scales. 
A series of measurements has been performed to investigate nuclear effects at 
high energies. For example, the European Muon Collaboration (EMC) had 
measured the ratio
of the electron-iron DIS cross sections to the electron-deuteron DIS cross section, which were found 
to be inconsistent with the ratios of the numbers of nucleons in iron and deuteron~\cite{EMC}. 
Later, similar measurements were performed by the Jefferson lab with more nuclei included, such as  
carbon, tritium, helium and beryllium, and gave similar conclusions~\cite{JLab1, JLab2}.
The inconsistency, known as the EMC effect, has become the most important experimental 
evidence that the high energy nuclear structure could be different from the simple picture of bonded nucleons 
with small correlations.
However, these experiments observe the relative difference between heavy nuclei and deuteron, 
whereas the analysis presented in this work probes the intrinsic (absolute) nuclear effect of 
the deuteron itself.
Besides, in the 
electron-nucleus DIS measurements, the physical interactions are dominated by the contributions of 
the valence quarks, i.e. $u$ and $d$, while this work focuses on the antiquarks. 
Corrections arising from potential nuclear effects may differ significantly between valence 
and sea quarks, as valence quark distributions dominate over sea quark contributions 
at $x\sim 0.1$. 
In conclusion, the discrepancy 
between the pure proton measurements and the deuteron data 
regarding the $\bar{d}/\bar{u}$ ratio may indicate a significant 
nuclear effect of the deuteron which has never been directly observed before. 
Furthermore, 
On the other hand, when compared to existing experimental results, this 
work suggested the potential presence of previously unrecognized nuclear effects 
in the deuteron. To develop a more comprehensive understanding of nuclear 
structure, new measurements, particularly those targeting valence quark-dominated DIS 
and sea quark-dominated Drell-Yan processes, are urgently needed. 

This work is also crucial for the study of nucleon structure. In modern global analysis 
of proton structure~\cite{CT18, MSHT, NNPDF}, a wide range of nuclear data, particularly 
from targets such as deuteron, iron, copper and lead, is used to provide constraints, especially 
in the large-$x$ region. 
As this article 
already demonstrates the potential inconsistency between nucleus and proton data, it would be 
important to perform the global analysis of the proton structure from pure proton interactions, such 
as the electron-proton DIS and proton-(anti)proton Drell-Yan, jet and top quark data. 
In the short term, a novel global analysis excluding nuclear data would provide a hint. However, 
in the long run, new measurements of proton interactions, such as the proton-proton collisions 
at the LHC, are needed to obtain (anti)quark-related information, allowing the replacement of 
legacy DIS data derived from nuclei.

~\\

\section{Global Analysis}
In the study of high-energy particle collisions, the internal structure of protons 
and neutrons, which is the fundamental building blocks of matter, is described 
by Parton Distribution Functions (PDFs). These functions quantify the probability 
of finding a parton (a quark or gluon) that carries a specific faction ($x$) of 
the parent hadron's momentum. The flavour structure of PDFs is defined 
by distinct contributions from various quark flavours (up, down, strange, charm, bottom) 
and gluons. This structure is critical for accurately modeling processes in 
high-energy physics, as different processes exhibit sensitivity to different 
combinations of these distributions. In this work, we closely adopt the 
methodology employed by the CTEQ-TEA collaboration for PDF 
extraction~\cite{CT18}. The core of the CTEQ-TEA PDF fitting 
procedure relies on a set of polynomial parameterization forms, 
which are initially postulated at an energy scale ($Q_0$) of approximately 
1 GeV. These parameterization are specifically designed to describe the PDFs 
of various parton flavours within the proton with high accuracy. At any arbitrary 
high-energy scale ($Q$), the PDFs are evolved using the Dokshitzer-Gribov-Lipatov-Altarelli-Parisi (DGLAP) 
evolution equations, leveraging the nonperturbative input PDFs determined at the $Q_0$ scale. 

In this work, the nonperturbative contributions to the PDFs of the (anti)quarks and gluons 
inside the proton are parameterized at $Q_0=1.3$ GeV. Below, we detail a nominal 
set of PDFs studied in this work. The key conclusion derived from this PDF set 
remains consistent across the various alternative fits we have explored. 
For the gluon distribution, the non-perturbative function is chosen to be:

\begin{footnotesize}
\begin{eqnarray}
 g(x) &=& a_0 x^{a_1 - 1} (1-x)^{a_2} P^g_a(y) \nonumber \\
 P^g_a(y) &=& \sinh[a_3](1-y)^3 + \sinh[a_4]3y(1-y)^2 \nonumber \\
   & & + a_5 3 y^2 (1-y) + y^3, 
\end{eqnarray}
\end{footnotesize}

\noindent where $y\equiv \sqrt{x}$, and $a_5$ is fixed as $a_5 = (3+2a_1)/3$.  
For valence up-quark ($u_V$) and down-quark ($d_V$), the non-perturbative 
functions are:

\begin{footnotesize}
\begin{eqnarray}\label{eq:valence}
  f_q(x) &=& a_0 x^{a_1 - 1}(1-x)^{a_2}P^V_a(y) \nonumber \\
    P^V_a(y) &=& \sinh[a_3](1-y)^4 + \sinh[a_4]4y(1-y)^3 \nonumber \\
     && + \sinh[a_5]6y^2(1-y)^2 + \left(1+\frac{1}{2}a_1 \right) 4y^3(1-y) \nonumber \\
     && + y^4,
\end{eqnarray}
\end{footnotesize}

\noindent where $y\equiv \sqrt{x}$. For both $u_V$ and $d_V$, $a_0$ is
fixed to yield $\int^1_0 u_V(x)dx = 2$ and $\int^1_0 d_V(x)dx = 1$, respectively, as the flavour 
number sum 
rules require. 
We further require $a^{u_V}_1 = a^{d_V}_1$ and
$a^{u_V}_2 = a^{d_V}_2$, as done in the CT18 global analysis, to ensure that the 
ratio between the two valence quarks take a finite value. 
For the sea quarks $\bar{u}$, $\bar{d}$, and the strangeness quark $s(=\bar{s})$, the
non-perturbative functions are:

\begin{footnotesize}
\begin{eqnarray}\label{eq:sea}
  f_q(x) &=& a_0 x^{a_1 - 1}(1-x)^{a_2}P^{\text{sea}}_a(y) \nonumber \\
    P^{\text{sea}}_a(y) &=&  (1-y)^5 + a_4 5y(1-y)^4 + a_5 10y^2(1-y)^3 \nonumber \\
    & & + a_6 10 y^3 (1-y)^2 + a_7 5y^4(1-y) + a_8y^5,
\end{eqnarray}
\end{footnotesize}

\noindent where $y\equiv 1- (1-\sqrt{x})^{a_3}$. 
In the CT18 global analysis, the shape parameters are required 
to have $a^{\bar{u}}_0 = a^{\bar{d}}_0$,
$a^{\bar{u}}_1 = a^{\bar{d}}_1 = a^{s}_1$, $a^{\bar{u}}_2 = a^{\bar{d}}_2$,
$a^{\bar{u}}_3 = a^{\bar{d}}_3 = a^{s}_3 = 4$, $a^{\bar{d}}_8=a^{s}_8 = 1$,
$a^{s}_4=a^s_5$, and $a^s_6=a^s_7$. Besides, $a^{\bar{u}}_0$ and $a^{\bar{d}}_0$ 
are directly given by the sum rule of $\sum_q \int^1_0 xq(x) \text{d} x=1$, after 
the normalization of other quarks are determined.
In this work, $a^{\bar{u}}_2$ and $a^{\bar{d}}_2$ are independent 
parameters, as the new measurements on $R$ can effectively constrain the 
difference between up and down flavours at $x\sim 0.1$. To fully consider the possible shift on 
antiquarks, $a^{\bar{d}}_0$ is set to be a free parameter in this work, while $a^{\bar{u}}_0$ 
is still calculated according to the momentum sum rule. $a^{\bar{d}}_1$, $a^{\bar{u}}_1$ and $a^{\bar{s}}_1$ 
are also set to be independent free parameters, as changes in the large $x$ region 
could possibly affect the small $x$ PDFs via the sum rules. 
In general, the choice of the non-perturbative functions used in this work is the same as
that used in the CT18 analysis, but with four more free parameters introduced. 

The non-perturbative functions, after being evolved to the corresponding energy scales 
of the input data using the DGLAP equations, 
provide the effective distributions of quarks and gluons, which 
are convoluted with the hard scattering cross sections to make predictions 
on the experimental observables of the data. 
The hard scattering calculations, following the same strategy as used in the 
CT18 global analysis, are generally performed at
the next-to-next-to-leading order (NNLO) in QCD interactions. The 32 free parameters in the non-perturbative functions
are determined by 
comparing the predicted values of experimental observables to the values measured in data, and 
requiring the minimal $\chi^2$ defined as:

\begin{footnotesize}
\begin{eqnarray}\label{eq:chi2}
\chi^2(a, \lambda) = \sum_k \frac{1}{s^2_k} \left( D_k - T_k(a) - \sum_\alpha \lambda_\alpha\beta_{k\alpha}  \right)^2 + \sum_\alpha \lambda^2_\alpha, 
\end{eqnarray}
\end{footnotesize}

\noindent where $D_k$ and $T_k(a)$ are the measured value and predicted value of the 
$k$-th observable included in the analysis. 
$s_k$ represents the uncertainty corresponding to $D_k$ and which is uncorrelated with the other data points. 
The correlated uncertainty between $D_\alpha$ and $D_k$ is represented by $\beta_{k\alpha}$, with a 
nuisance parameter $\lambda_\alpha$ introduced in the analysis. Details of the establishment of 
Eq.~\eqref{eq:chi2} is discussed in Sect. III-A of Ref.~\cite{CT18}.

The uncertainty of the global analysis is estimated using the Hessian method. In such method, the 
variation of $\chi^2$ as a function of the fitted parameters $a$ is represented in forms 
of a set of orthogonal eigenvectors, of which each corresponds to an independent direction in 
the space of the fitted parameters. For a typical $\chi^2$ defined as in Eq.~\eqref{eq:chi2}, the 
uncertainty can be quoted as the difference between the best fitted parameter values $a_0$ and 
the shifted parameter values $a'$, corresponding to $\Delta \chi^2 = \chi^2(a') - \chi^2_\text{min}(a_0) = 1$. 
In CT18, the uncertainty was enlarged by $\Delta\chi^2\approx 37$ to cover potential biases 
in the global analysis, 
such as the arbitrariness of the non-perturbative function formality, and tensions between experimental 
results. Particularly, the potential differences in the underlying physics of 
the experimental measurements, such as the unknown nuclear effect being studied in this work, 
are ignored. 
In this work, we consistently use the criterion $\Delta \chi^2=1$ to  
estimate the uncertainties associated with the determined quark and gluon distributions, which 
represents another difference in the analysis strategy relative to the CT18.  
The reason for using the original uncertainty definition of $\Delta \chi^2 = 1$ in this work is to avoid 
obscuring difference in the underlying physics between the pure proton datasets and 
the deuteron datasets.

\begin{figure*}[!h]
\begin{center}
\begin{subfigure}{0.45\textwidth}
\includegraphics[width=\textwidth]{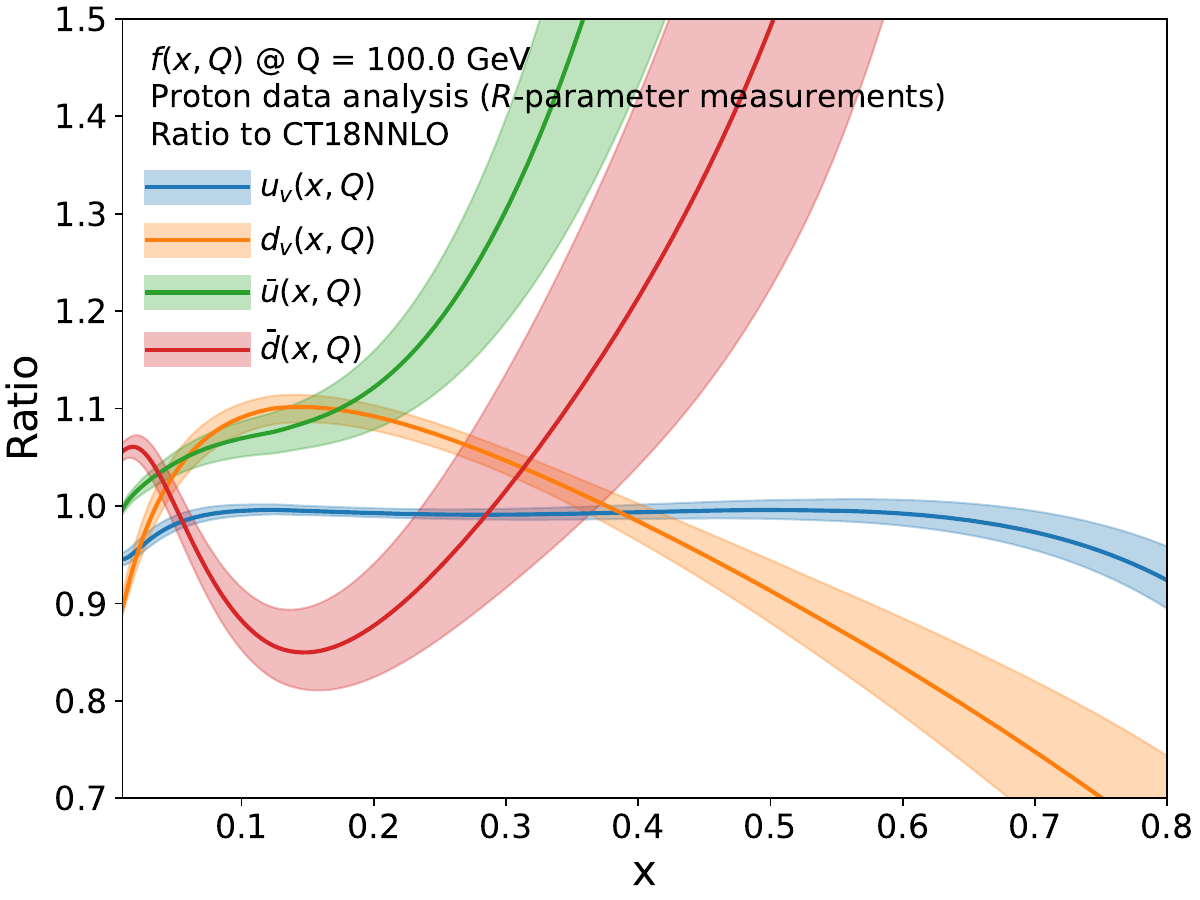}
\caption{}
\end{subfigure}
\begin{subfigure}{0.45\textwidth}
\includegraphics[width=\textwidth]{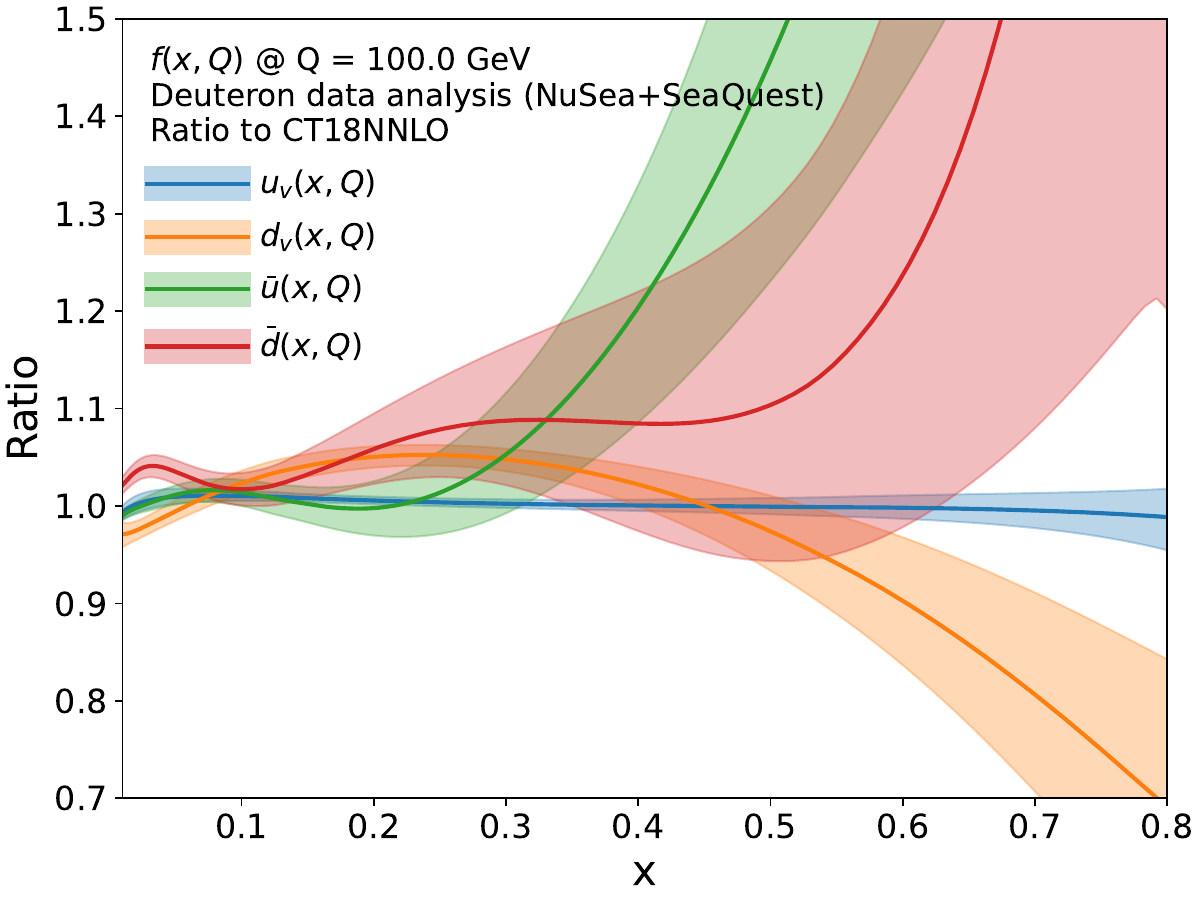}
\caption{}
\end{subfigure}
\begin{subfigure}{0.45\textwidth}
\includegraphics[width=\textwidth]{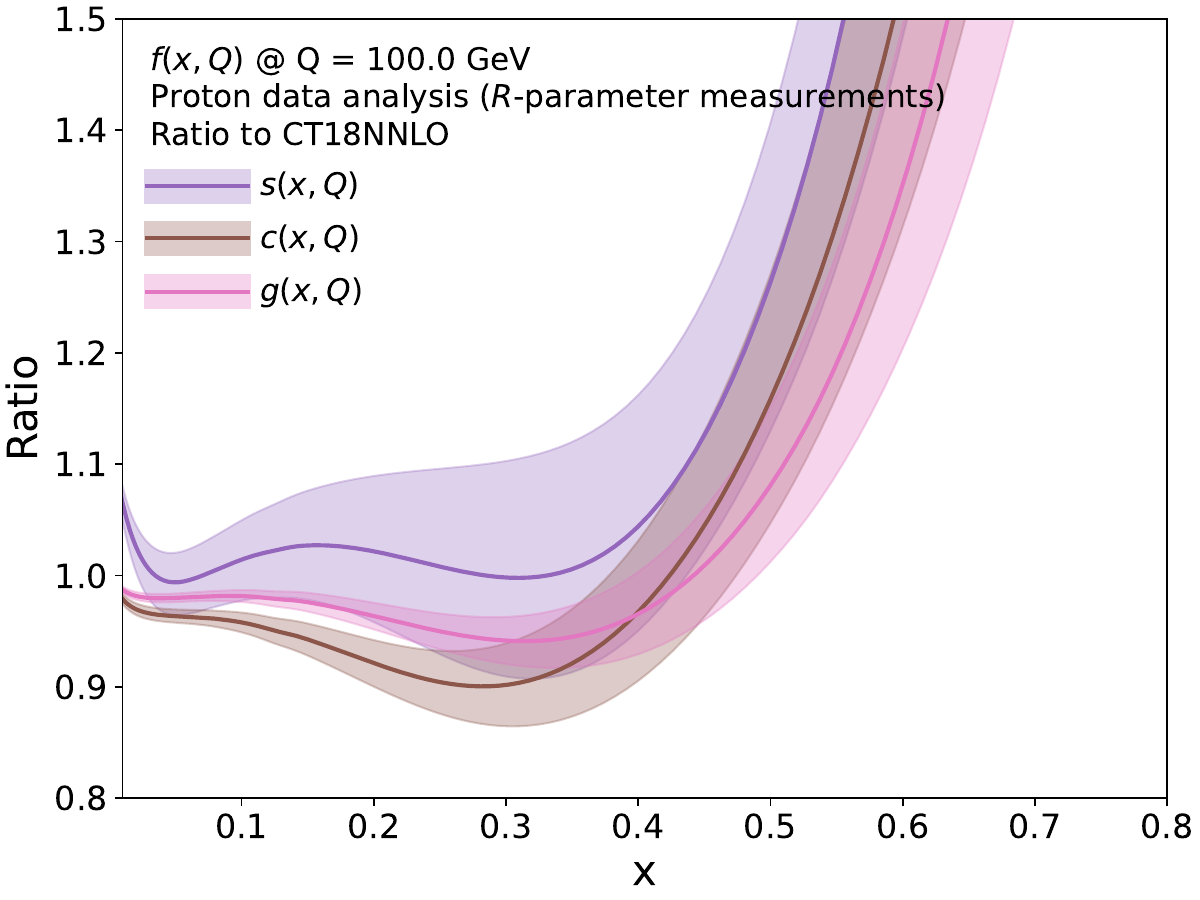}
\caption{}
\end{subfigure}
\begin{subfigure}{0.45\textwidth}
\includegraphics[width=\textwidth]{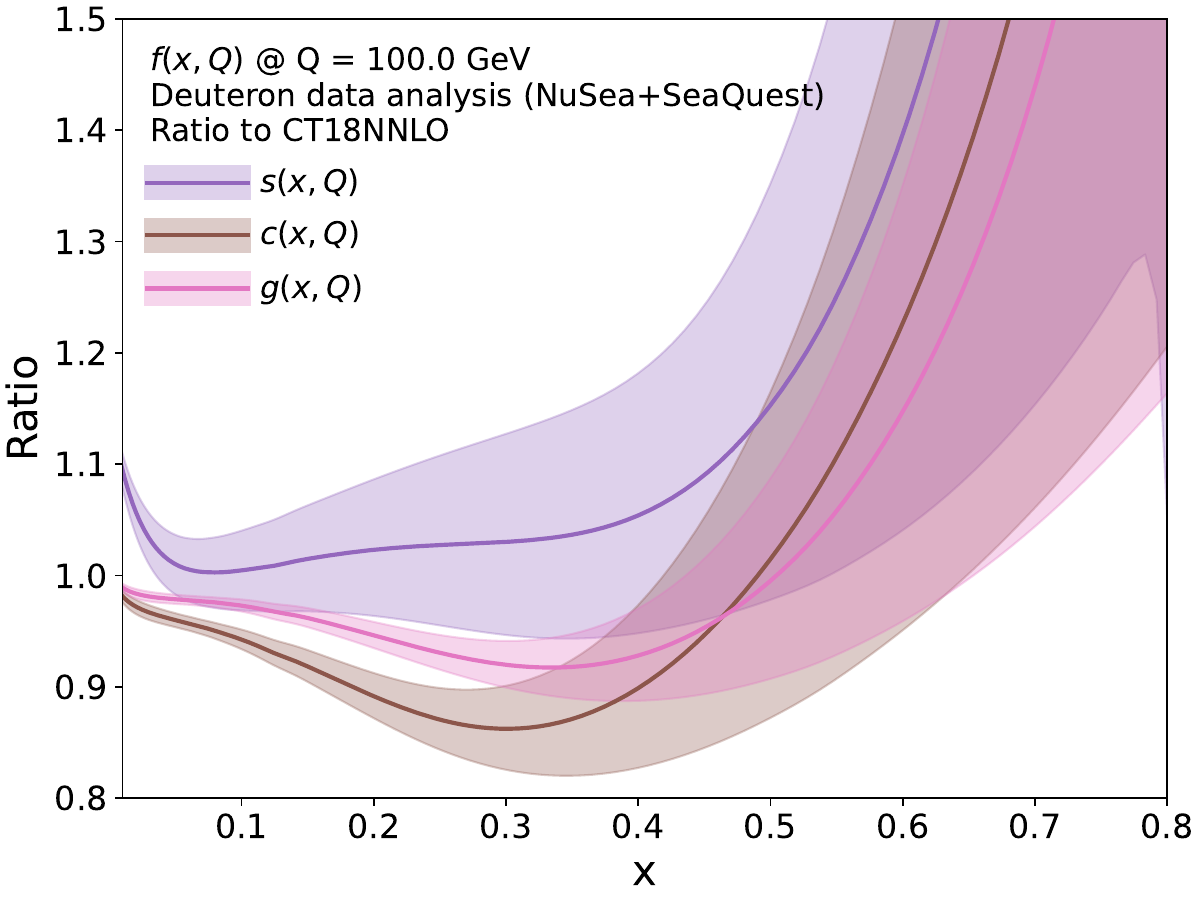}
\caption{}
\end{subfigure}
\caption{\small {\bf Relative distributions of quarks and gluon.} 
The quark and gluon distributions derived from the analysis of the pure proton dataset and 
that of the deuteron dataset, demonstrated as the ratio with respect to the corresponding predictions 
from CT18NNLO PDF. For the distributions of $u(x)$, $d(x)$, $\bar{u}$ and $\bar{d}$, the ratios 
are given in (a) and (b) for the proton data and the deuteron data analysis, respectively. 
For the distributions of $s(x)=\bar{s}(x)$, $g(x)$ and $c(x)$, the ratios are given in 
(c) and (d) for the proton data and the deuteron data analysis, respectively. }
\label{fig:pdfs}
\end{center}
\end{figure*}

The main conclusion from this work, the ratio of $\bar{d}(x)/\bar{u}(x)$, is already demonstrated in Fig.~\ref{fig:flavorasym}. 
To provide additional context for the analysis, we also present, 
in Fig.~\ref{fig:pdfs}, the ratios 
of the distributions of other quark flavours and the gluon to the corresponding predictions 
from the CT18NNLO PDF sets. 
For $u(x)$, $d(x)$, $\bar{u}(x)$ and $\bar{d}(x)$, the deuteron data analysis 
yields results that are more consistent with the CT18 predictions. This consistency arises because the deuteron data of 
NuSea was included as a key constraint in the CT18 analysis. 
Notably, in the proton data analysis, our results yield distributions that differ substantially from 
those the of CT18. This is expected, as the novel measurements of the $R$ parameters 
(which we incorporate here) were not available at the time of the CT18 analysis. 
For $s(x)=\bar{s}(x)$ and $g(x)$, the two analyses in this work 
yield similar distributions, as neither the $R$-parameter measurements nor the deuteron data 
provides sufficient constraints on these flavours. 
Instead, they are primarily determined by other experimental data included in the PDF 
global analysis. Furthermore, the distribution of charm quark, $c(x)=\bar{c}(x)$, is also 
consistent between the two analyses, as
the charm quarks are taken to be perturbatively generated in this analysis.
~\\

~\\

\section{Acknowledgement}
This work was supported by the National Natural Science Foundation of China under Grant No. 12061141005. 
This work was also supported by the U. S. National Science Foundation under Grant No. PHY-2310291. We thank Dr. Yang Li from 
the University of Science and Technology of China and our CTEQ-TEA colleagues for many helpful discussions.
~\\

\end{document}